\documentclass[ALICE,manyauthors]{cernphprep}

\usepackage[comma,square,numbers,sort&compress]{natbib}
\usepackage{hyperref}
\usepackage{lineno}
\usepackage{dcolumn}
\usepackage{bm}
\usepackage{amsmath}
\usepackage{multirow}
\usepackage{relsize}
\usepackage{environ}
\usepackage{placeins}
\usepackage{amssymb}
\usepackage{float}
\usepackage{xspace}
\usepackage{needspace}
\usepackage{color,soul}

\newcommand{\nudyn}{$\nu_{\rm dyn}$\xspace}

\newcommand{\nudynpikashort}{$\nu_{\rm{dyn}}[\pi,\mathrm{K}]$\xspace}
\newcommand{\nudynpiprshort}{$\nu_{\rm{dyn}}[\pi,\mathrm{p}]$\xspace}
\newcommand{\nudynkaprshort}{$\nu_{\rm{dyn}}[\mathrm{p},\mathrm{K}]$\xspace}
\newcommand{\pt}{\ensuremath{p_{\rm T}}\xspace}
\newcommand{\dEdx}{d$E$/d$x$\xspace}

\newcommand{\dNdeta}{d$N_{\rm ch}/$d$\eta$\xspace}

\newcommand{\snn}{$\sqrt{s_{\rm{NN}}}$\xspace}

\newcommand{\pp}{\ensuremath{\mathrm {p\kern-0.05em p}}\xspace}
\newcommand{\PbPb}{\ensuremath{\mbox{Pb--Pb}}\xspace}

\newcommand{\eqnref}[1]{Eq.~\ref{#1}}
\newcommand{\figref}[1]{Figure~\ref{#1}}
\newcommand{\Tabref}[1]{Table~\ref{#1}}

\newcommand{\Secref}[1]{Section~\ref{#1}}
\newcolumntype{C}[1]{>{\centering\arraybackslash}m{#1}}
\sloppypar

\begin{document}

\begin{titlepage}
\PHyear{2017}
\PHnumber{318}      
\PHdate{1 December}  
\title{Relative particle yield fluctuations \\in \PbPb collisions at \mbox{\snn$=2.76$~TeV}}
\ShortTitle{Relative particle yield fluctuations in \PbPb collisions}   
\Collaboration{ALICE Collaboration%
  \thanks{See Appendix~\ref{app:collab} for the list of collaboration members}}
\ShortAuthor{ALICE Collaboration} 
\begin{abstract}
First results on K/$\pi$, p/$\pi$ and K/p fluctuations are obtained with the ALICE detector at the CERN LHC as a function 
of centrality in \PbPb collisions at \mbox{\snn$=2.76$~TeV}. The observable \nudyn, which is defined in terms of the moments 
of particle multiplicity distributions, is used to quantify the magnitude of dynamical fluctuations of relative particle 
yields and also provides insight into the correlation between particle pairs. This study is based on a novel experimental 
technique, called the Identity Method, which allows one to measure the moments of multiplicity distributions in case of incomplete 
particle identification. The results for p/$\pi$ show a change of sign in \nudyn from positive to negative towards more peripheral 
collisions. For central collisions, the results follow the smooth trend of the data at lower energies and \nudyn exhibits a change in 
sign for p/$\pi$ and K/p.
\end{abstract}
\end{titlepage}
\setcounter{page}{2}

\section{\label{sec:intro}Introduction}

The theory of strong interactions, Quantum Chromodynamics (QCD), predicts that at sufficiently high energy density 
nuclear matter transforms into a deconfined state of quarks and gluons known as Quark-Gluon Plasma (QGP) \cite{Collins:1974ky,Shuryak:1980tp}. 
One of the possible signatures of a transition between the hadronic and partonic phases is the enhancement of fluctuations of the number 
of produced particles in the hadronic final state of relativistic heavy-ion collisions \cite{Stephanov:1998dy,Shuryak:2000pd,Bazavov:2012jq}. 
Event-by-event fluctuations and correlations may show critical behaviour near the phase boundary, including the crossover region where 
there is no thermal singularity, in a strict sense, associated with the transition from a QGP phase to a hadron-gas phase. A correlation 
analysis of event-by-event abundances of pions, kaons and protons produced in \PbPb collisions at LHC energies may provide a connection 
to fluctuations of globally conserved quantities such as electric charge, strangeness and baryon number, and therefore shed light on 
the phase structure of strongly interacting matter \cite{Koch:2008ia}. 
\\ \\ 
In view of the predicted criticality signals at crossover for vanishing net-baryon densities \cite{Bazavov:2011nk}, 
event-by-event fluctuations of relative particle yields are studied using the fluctuation measure $\nu_{\rm dyn}[A,B]$ \cite{Pruneau:2002yf} 
defined in terms of moments of particle multiplicity distributions as 
\begin{equation} \label{eq:nudyn}
  \nu_{\rm dyn}[A,B] = \dfrac{\langle N_{A}(N_{A}-1) \rangle}{{\langle N_{A} \rangle}^{2}} + 
  \dfrac{\langle N_{B}(N_{B}-1) \rangle}{{\langle N_{B} \rangle}^{2}} - 
    2\dfrac{\langle N_{A}N_{B} \rangle}{\langle N_{A} \rangle\langle N_{B} \rangle},
\end{equation}
\noindent 
where $N_{A}$ and $N_{B}$ are the multiplicities of particles $A$ and $B$ measured event-by-event in a given kinematic range. 
The $\nu_{\rm dyn}[A,B]$\footnote{In this study, $\nu_{\rm dyn}[A,B]$ was taken to be $\nu_{\rm dyn}[A+\overline{A},B+\overline{B}]$, where 
  $\overline{A}$ and $\overline{B}$ are the anti-particles of $A$ and $B$, respectively.} 
fluctuation measure contrasts the relative strength of 
fluctuations of species $A$ and $B$ to the relative strength of correlations between these two species.
It vanishes when the particles $A$ and $B$ are produced in a statistically independent way \cite{Pruneau:2002yf,Christiansen:2009km}. 
\\ \\ 
This study at LHC energies is of particular importance for establishing the energy and system size dependence of \nudyn in order to 
understand the trend observed at lower collision energies from the RHIC Beam Energy Scan (BES) results reported by the STAR collaboration \cite{Abdelwahab:2014yha}. 
Furthermore, the advantage of this fluctuation 
measurement is its robustness against non-dynamical contributions such as those stemming from participant nucleon fluctuations and finite 
particle detection efficiencies \cite{Pruneau:2002yf, Braun-Munzinger:2016yjz}. 
Measurements of the \nudyn observable for net-charge fluctuations were already published by ALICE \cite{Abelev:2012pv}. Moreover, for identified particles, 
it was measured at the Super Proton Synchrotron (SPS) \cite{Anticic:2013htn} and at the Relativistic Heavy-Ion Collider (RHIC) \cite{Abdelwahab:2014yha} in \PbPb 
and Au--Au collisions, respectively.
The ALICE detector at the LHC is ideally suited to extend these measurements to higher collision energies. In particular, the excellent 
charged-particle tracking and particle identification (PID) capabilities in the central barrel of the detector allow for a 
precise and differential event-by-event analysis at midrapidity and low transverse momentum (\pt). 
\\ \\
The paper is organized as follows. In \Secref{sec:alice}, details about the ALICE detector setup and the dataset are given. 
\Secref{sec:dataana} discusses the event and track selection criteria, particle identification procedure, and the 
analysis method. Estimates of statistical and systematic uncertainties are given in \Secref{sec:error}. 
Results on \mbox{\nudynpikashort}, \mbox{\nudynpiprshort} and \mbox{\nudynkaprshort} in \PbPb collisions at 
\mbox{\snn$=2.76$~TeV} are presented in \Secref{sec:results}, and finally \Secref{sec:summary} summarizes the measurements presented in this paper.
%
%
\section{\label{sec:alice}Experimental setup and dataset}
ALICE is a general-purpose detector system designed, in particular, for the study of collisions of heavy ions at the LHC. 
The design, components, and performance of the ALICE detector have been reported elsewhere \cite{Abelev:2014ffa, Aamodt:2008zz}. The ALICE detector is 
comprised of several detector components organized into a central barrel detection system and forward/backward detectors. 
The main tracking and PID devices 
in the central barrel of the experiment are the Inner Tracking System (ITS) and the Time Projection Chamber (TPC), which are operated
inside a large solenoidal magnet with $B=0.5$\,T. 
Two forward scintillator arrays V0-A and V0-C are located on either side of the interaction point and cover the pseudorapidity ($\eta$) 
intervals $2.8<\eta<5.1$ and $-3.7<\eta<-1.7$. The V0 detectors and the two neutron Zero Degree Calorimeters (ZDC), placed at $\pm$114\,m from 
the interaction point, were used for triggering and event selection.
\\ \\
The ITS-TPC tracking system  covers the midrapidity region and provides charged-particle tracking and momentum 
reconstruction down to $\pt=100$\,MeV/$c$. The ITS is employed to reconstruct the collision vertex with high precision and 
to reject charged particles produced in secondary vertices. 
\\ \\
The analysis presented in this paper is based on about 13 million minimum-bias \PbPb collisions at \mbox{\snn$=2.76$~TeV} collected in the year 2010.
The minimum-bias trigger condition is defined by the coincidence of hits in both V0 detectors. In the offline event selection, V0 and ZDC 
timing information is used to reject beam-gas background and parasitic beam-beam interactions.  
The definition of the collision centrality is based on the charged-particle multiplicity measured in the V0 detectors \cite{Abelev:2014ffa}, 
which can be related to collision geometry and the number of participating 
nucleons through a Monte-Carlo (MC) simulation based on a Glauber model \cite{Alver:2008aq}. 
%
%
\section{\label{sec:dataana}Data analysis}
\subsection{Event and track selection}
Charged particles reconstructed in the TPC with full azimuthal acceptance and in the pseudorapidity range of $|\eta|<$0.8 were used in this 
analysis. The momentum range was restricted to $0.2<p<1.5$~GeV/$c$ in order to minimize systematic uncertainties 
arising from the overlap of the \dEdx distributions.
Furthermore, the following track selection criteria were applied to guarantee optimal \dEdx and momentum resolution, which are crucial 
for precise particle identification. Charged-particle tracks were accepted in this analysis when they have at least 80 out of a maximum of 159 reconstructed 
space points in the TPC, and the $\chi^{2}$ per space point from the track fit is less than 4. Daughter tracks from 
reconstructed secondary weak-decay kink topologies were rejected. Additional suppression of secondary particles was achieved by restricting 
the distance-of-closest-approach (DCA) of the extrapolated trajectory to the primary vertex position to less than 2~cm 
along the beam direction. In the transverse plane the restriction in the DCA depends on \pt in order to take into account 
the \pt dependence of the impact parameter resolution \cite{Aamodt:2010jd}. The remaining contamination after the DCA cuts is typically less 
than 10\% for the momentum range covered in this work \cite{Abelev:2013vea}.
\subsection{Identity Method}
\begin{figure}[h]
  \centering
  \includegraphics[width=9cm]{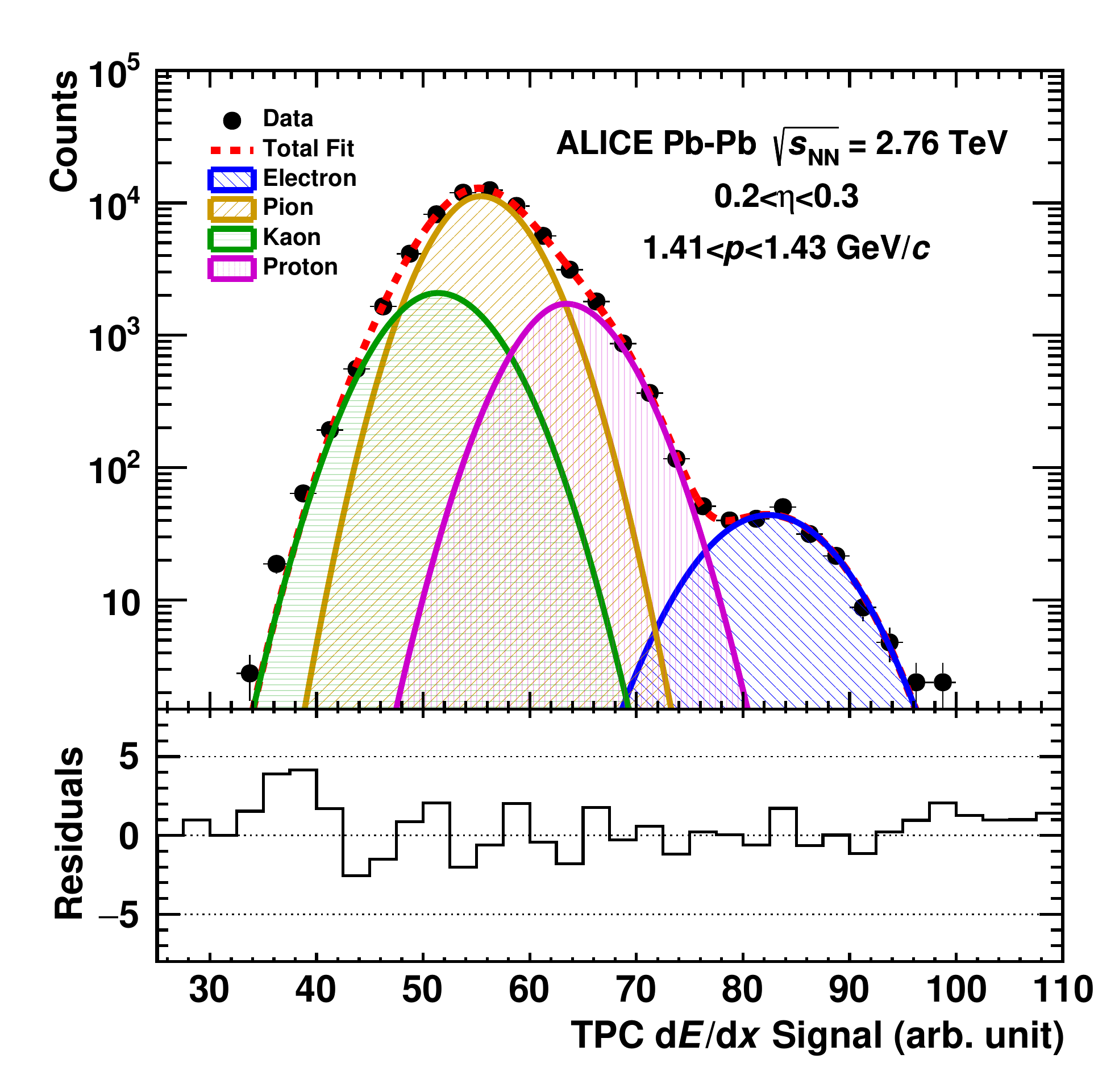}
  \caption{Distributions of the TPC \dEdx signal of pions, kaons, electrons and protons fitted with the generalized Gaussian function in a given phase-space bin. 
  The residuals are defined as the difference between data points and the total fit function normalized to the statistical error of the data points.}
\label{fig:dedxdist}
\end{figure}

The standard approach of finding the moments $\langle N_{A} \rangle$, $\langle N_{B} \rangle$, $\langle N_{A}(N_{A}-1) \rangle$ and 
$\langle N_{B}(N_{B}-1) \rangle$ is to count the number of particles $N_{A}$ and $N_{B}$ event-by-event and calculate averages over the dataset. 
However, this approach suffers from incomplete particle identification due to overlapping \dEdx distribution functions, which could be circumvented 
by either selecting suitable phase-space regions or by using additional detector information such as time-of-flight measurements. 
These procedures reduce the overall phase-space coverage and detection efficiencies. The present study is based on the Identity 
Method \cite{Gazdzicki:2011xz,Gorenstein:2011hr,Rustamov:2012bx} which overcomes the misidentification problem.
\\ \\
\begin{figure}[h]
  \centering
  \includegraphics[width=12cm]{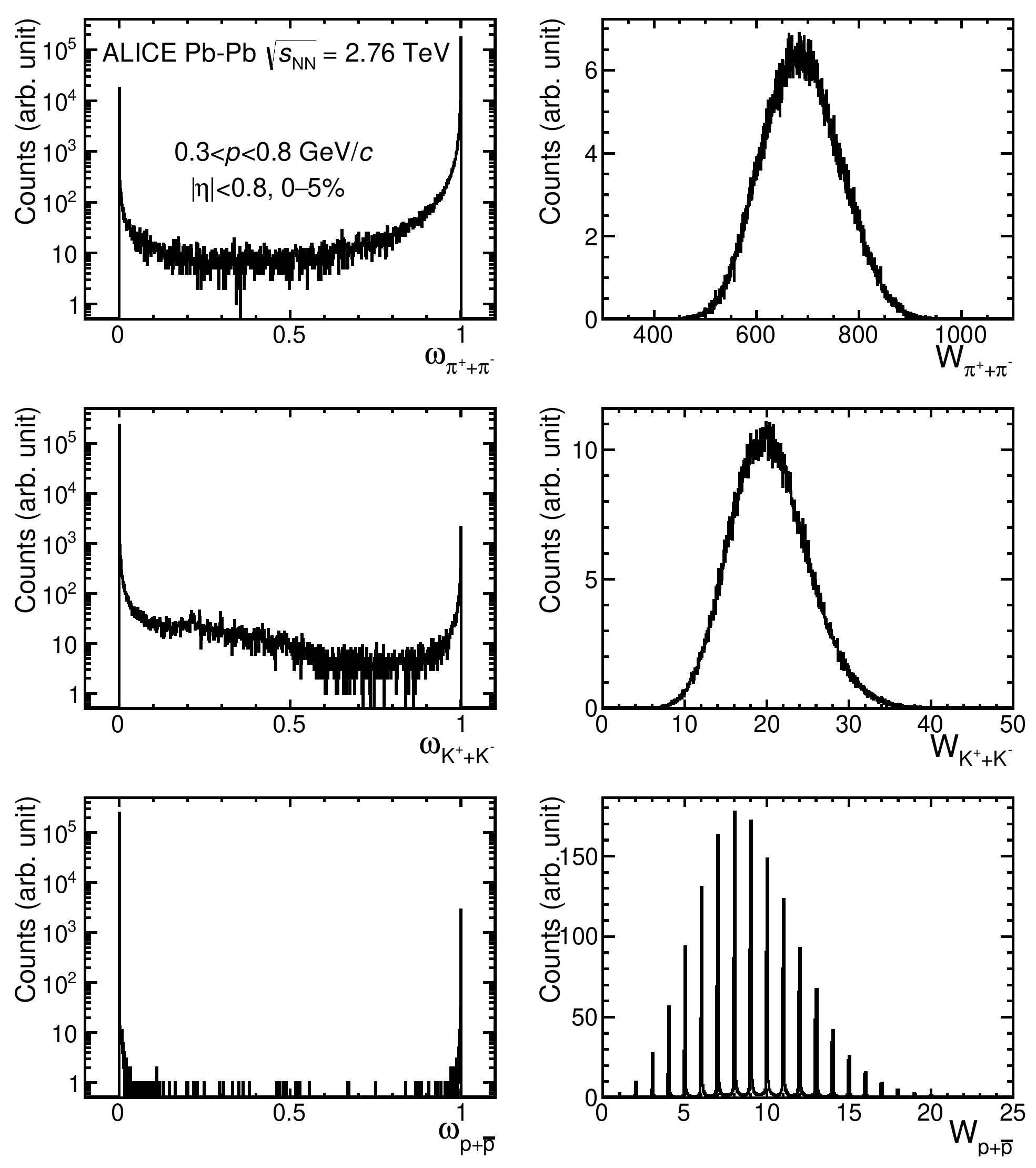}
  \caption{Distributions of $\omega$ and $W$ for pions (top), kaons (middle) and protons (bottom) in the momentum interval of $0.3<p<0.8$~GeV/$c$ for 
    0--5\% central \PbPb events.}
  \label{fig:wdist}
\end{figure}
The Identity Method was proposed in Ref.~\cite{Gazdzicki:2011xz} as a solution to the misidentification 
problem for the analysis of events with two different particle species. In Ref.~\cite{Gorenstein:2011hr}, the method was developed further 
to calculate the second moments of the multiplicity distributions of more than two particle species. Subsequently, in Ref.~\cite{Rustamov:2012bx}, 
it was generalized to the first and higher moments of the multiplicity distributions for an arbitrary number of particle species. 
The first experimental results using the Identity Method were published by the NA49 collaboration \cite{Anticic:2013htn}. 
\\ \\
Instead of counting every detected particle event-by-event, the Identity Method follows a probabilistic approach using 
two basic experimental per-track and per-event observables, $\omega$ and $W$, respectively. They are defined as 
\begin{equation} \label{eq:omega}
  \omega_{j}(x_{i}) = \dfrac{\rho_{j}(x_{i})}{\rho(x_{i})} \in[0,1],  \quad 
  \rho(x_{i}) = \sum_{j}\rho_{j}(x_{i}), 
  \quad W_{j} \equiv \sum_{i=1}^{N(n)} \omega_{j}(x_{i}),
\end{equation}
\noindent where $x_{i}$ stands for the \dEdx of a given track $i$, 
$\rho_{j}(x)$ is the \dEdx distribution of particle species $j$ within a given phase-space bin and $N(n)$ is the number of tracks in 
the $n^{\rm th}$ event. The quantity $\omega_{j}(x_{i})$ represents the probability that particle $i$ is of type $j$. 
Thus, in case of perfect particle identification, one expects $W_{j} = N_{j}$, while this does not hold in case of 
overlapping \dEdx distributions. \figref{fig:wdist} shows the $\omega$ and $W$ distributions for pions, kaons and protons in the momentum interval 
of $0.3<p<0.8$~GeV/$c$. The $W$ distribution of protons shows a discrete structure because proton \dEdx distributions have 
the least overlap.
\\ \\
The moments of the $W$ distributions can be constructed directly from experimental data. 
The Identity Method calculates the moments of the particle multiplicity distributions by unfolding the moments of the $W$ 
distributions with the following matrix operation
\begin{equation} \label{eq:matrixoper}
  \langle \overrightarrow{N} \rangle = \textrm{A}^{-1} \langle \overrightarrow{W} \rangle, 
\end{equation}
\noindent where $\langle \overrightarrow{W} \rangle$ and $\langle \overrightarrow{N} \rangle$ are the vectors of the moments of $W$ 
quantities and unknown true multiplicity distributions, respectively. The response matrix $\textrm{A}$ is defined by the $\omega$ quantities. 
A detailed description of the technique and a demonstration of its robustness can be found in Refs~\cite{Gorenstein:2011hr, Rustamov:2012bx}.
\\ \\ 
The \dEdx measurements used as the only input for the Identity Method are obtained from the TPC, which provides a 
momentum resolution of better than 2\% and a single-particle detection efficiency of up to 80\% for the kinematic range considered in this 
paper \cite{Abelev:2014ffa}.
The Identity Method employs fits of inclusive \dEdx distributions for the calculation of the $\omega$ 
probabilities entering \eqnref{eq:matrixoper}. 
Since the overlap regions in the \dEdx distributions are also properly taken into account, a very good description of the 
inclusive \dEdx spectra, and therefore an excellent understanding of the TPC detector response, is required over the full 
momentum range covered in this analysis.
To this end, the \dEdx distributions of pre-selected samples of pions, protons and electrons, identified by the reconstruction 
of $\mathrm{K}_{S}^{0}$ and $\varLambda$ decays and photon conversions, were fitted with a generalized Gaussian function of the form:
\begin{equation} \label{eq:gengaussfunc}
  f(x) = A e^{- {\left( |x-\mu|/\sigma \right)}^{\beta}} {\left( 1+\textrm{erf} \left( \alpha \dfrac{|x-\mu|}{\sigma \sqrt{2}} \right) \right)}
\end{equation}
where $A$, $\mu$, $\sigma$, $\alpha$ and $\beta$ stand for the abundance, mean, width, skewness and kurtosis of the distribution, respectively.
The detector response functions obtained in this way were used later to fit the inclusive \dEdx spectra. 
To cope with the dependencies of the \dEdx on the track angle and particle multiplicity, fits were 
performed over the entire pseudorapidity range of $|\eta|<0.8$ in steps of 0.1 units for each centrality class. 
Moreover, the momentum intervals were chosen narrow enough to minimize the effect of the momentum dependence on \dEdx, 
most particularly at low momenta where the magnitude of \dEdx varies rapidly with the momentum.
An example of a \dEdx distribution in a given phase-space bin and the corresponding fits are shown in Fig.~\ref{fig:dedxdist}.
%
%
\section{\label{sec:error}Statistical and systematic uncertainties}
The statistical uncertainties were determined by the number of events in this analysis and the finite number of tracks in each event. 
The number of events also affects the uncertainty of the shape of the inclusive \dEdx spectra, which is determined by a fit. 
This uncertainty enters the calculation of $\omega$ and $W$, and finally the computation of the moments 
of multiplicity distributions with the Identity Method. Since standard error propagation is impractical given the rather complicated numerical derivation of the final 
result, the subsample method was chosen to evaluate the statistical uncertainties. 
To this end, the data set was subdivided into $n=$~25 random subsamples $i$. 
The \nudyn values were reconstructed for each subsample and the statistical uncertainty was obtained according to
\begin{equation}
  \sigma_{\langle \nu_{\mathrm{dyn}} \rangle} = \dfrac{\sigma}{\sqrt{n}},
\end{equation}
where 
\begin{equation}
  \sigma = \sqrt{\dfrac{\sum_{i} (\nu_{\mathrm{dyn},i} - \langle \nu_{\mathrm{dyn}} \rangle )^2}{n-1}}, \quad  \quad 
  \langle \nu_{\mathrm{dyn}} \rangle = \dfrac{1}{n}\sum_{i} \nu_{\mathrm{dyn},i}.
\end{equation}
The summary of all sources of systematic uncertainties is shown in \Tabref{tab:systerrors} and in the next paragraphs the main contributors 
to the systematics are detailed. 
\\ \\
The largest contribution to the total systematic uncertainty is from the fits of the measured particle \dEdx distributions. 
The quality of the fits was monitored by Kolmogorov-Smirnov (K-S) and $\chi^{2}$ tests.
To study the influence of possible systematic shifts in the fit parameters on \nudyn,
the fit parameters of each particle in the overlap regions were varied by about $\pm$0.5~\%, which defines the boundaries where the K-S test fails at 90\% 
confidence level. 
The observed maximum variations range from about 7\% to 15\% for \nudynpiprshort and \nudynpikashort, respectively.
\\ \\
Even though \nudyn is known to be robust against detection-efficiency losses, it may show an explicit dependence if the detector 
response functions differ from Binomial or the efficiencies exhibit large variations with detector occupancy 
\cite{Pruneau:2002yf}. Therefore, one also has to investigate 
the uncertainty resulting from the detection efficiency losses. For that, the \nudyn results reconstructed from a full Monte Carlo simulation 
of HIJING \cite{Gyulassy:1994ew,Deng:2010mv} events employing a GEANT3 \cite{Brun:1994aa} implementation of the ALICE detector were compared to the analysis at 
the generator level, where in both generated and reconstructed levels perfect PID information was used. The resulting systematic uncertainty from the 
finite tracking efficiency is less than 6\%.
\begin{table}
  \centering
  \caption{List of contributions to the systematic uncertainty of the particle ratio fluctuations.}
  \scalebox{1}{
    \begin{tabular}{  l| c | c | c }
      \hline \hline
      Uncertainty source          & \nudynpikashort~(\%) & \nudynpiprshort~(\%) & \nudynkaprshort~(\%)  \\ 
      \hline
      Inclusive \dEdx fits        & 10--15    & 4--7      & 8--12       \\
      Detection efficiency        & 0.5--6    & 0.5--4      & 0.5--5       \\
      DCA to vertex               & 1--4      & 1--2      & 1--3       \\
      Vertex $z$ position         & 0.5--2    & 0.5--1    & 0.5--2       \\
      TPC $\chi^{2}/d.o.f.$       & 1--3      & 1--2      & 1--3       \\
      Min. TPC space points       & 0.5--3    & 0.5--2    & 0.5--3       \\
      B-field polarity            & 0.5--2    & 0.5--1    & 0.5--2       \\
      \hline
      Total systematic uncertainty      & 10--17    & 4--9      & 8--14       \\
      \hline \hline
    \end{tabular}
  }
  \label{tab:systerrors}
\end{table}
\\ \\
The systematic uncertainties due to the track selection criteria were estimated by 
a variation of the selection ranges. The systematics from contamination of weak decays and other secondary particles 
were obtained by varying the DCA cuts.
Other contributions to the total systematic uncertainty arise from the cuts applied on the maximum 
distance of the reconstructed vertex to the nominal interaction point along the beam axis, the number of required TPC 
space points per track and the $\chi^{2}$ per degree of freedom of the track fit.
Moreover, the effect of the magnetic field polarity was investigated by separate analyses of data taken under two polarities. 
Neither of these contributions to the total systematic uncertainty exceeds 5\%.
The total systematic uncertainty was obtained by adding in quadrature the individual 
maximum systematic variations from these different contributions.
%
%
\section{\label{sec:results}Results}
\subsection{Centrality dependence and comparison to models}
\begin{figure}[h]
  \centering
  \includegraphics[width=8.5cm]{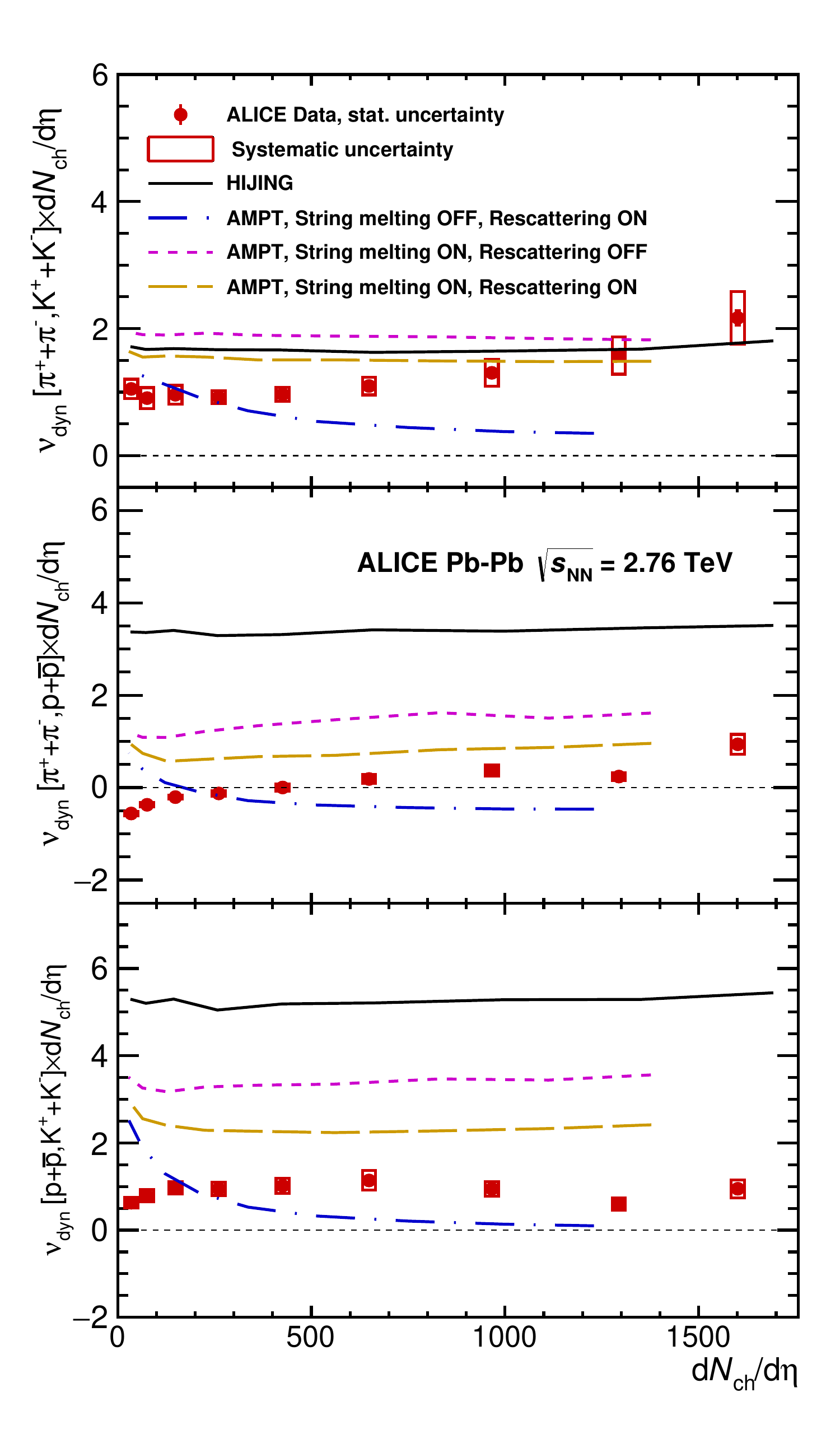}
  \caption{
    Results for \nudynpikashort, \nudynpiprshort and \nudynkaprshort scaled by the charged-particle density \dNdeta. The ALICE data are shown by red 
    markers while the coloured lines indicate the HIJING \cite{Gyulassy:1994ew,Deng:2010mv} and AMPT \cite{Lin:2004en} model calculations. 
    The data are shown as a function of the collision centrality, expressed in terms of \dNdeta. 
  }
  \label{fig:centralitydep}
\end{figure}

In this section, the results are presented as a function of collision centrality and compared to calculations with the HIJING \cite{Gyulassy:1994ew,Deng:2010mv} 
and AMPT \cite{Lin:2004en} models. The unscaled values of \nudyn for different combinations of particles in each centrality class, together with 
the final statistical and systematic uncertainties, are given in \Tabref{tab:nudynresults}. Due to the intrinsic multiplicity dependence of \nudyn, 
discussed in Refs~\cite{Koch:2009dg,Abelev:2009ai}, the values of \nudyn were scaled 
further by the charged-particle multiplicity density at midrapidity, \dNdeta. The fully corrected experimental \dNdeta values were taken 
from Ref~\cite{Abelev:2013vea}. \figref{fig:centralitydep} shows measured values of \nudyn scaled by \dNdeta as a function of the collision centrality expressed 
in terms of \dNdeta. The values for \nudyn and \dNdeta for HIJING and AMPT were calculated by using corresponding particle multiplicities at the 
generator level within the same experimental acceptance. A flat behaviour is expected in this representation if a superposition of 
independent particle sources is assumed, as in the Wounded Nucleon Model (WNM) \cite{Bialas:1976ed}.
\begin{table}[h]
  \centering
  \caption{Numerical values of \nudyn results for different particle pairs. The first uncertainty is statistical and the second systematic.}
  \scalebox{1}{
    \begin{tabular}{  c | c | c | c | c  }
      \hline \hline
      Centrality (\%)      & $\langle$\dNdeta$\rangle$   & \nudynpikashort~($10^{-3}$)    & \nudynpiprshort~($10^{-3}$)    & \nudynkaprshort~($10^{-3}$) \\
      \hline
      0--5                 & 1601$\pm$60                 & 1.35  $\pm$0.08  $\pm$0.25     & 0.59    $\pm$0.08  $\pm$0.13   & 0.59   $\pm$0.08  $\pm$0.13 \\
      5--10                & 1294$\pm$49                 & 1.22  $\pm$0.08  $\pm$0.22     & 0.19    $\pm$0.08  $\pm$0.06   & 0.46   $\pm$0.10  $\pm$0.11 \\
      10--20               & 966$\pm$37                  & 1.35  $\pm$0.08  $\pm$0.21     & 0.38    $\pm$0.08  $\pm$0.12   & 0.98   $\pm$0.10  $\pm$0.17 \\
      20--30               & 649$\pm$23                  & 1.69  $\pm$0.09  $\pm$0.21     & 0.29    $\pm$0.09  $\pm$0.15   & 1.76   $\pm$0.13  $\pm$0.34 \\
      30--40               & 426$\pm$15                  & 2.27  $\pm$0.11  $\pm$0.25     & 0.01    $\pm$0.18  $\pm$0.18   & 2.39   $\pm$0.24  $\pm$0.40 \\
      40--50               & 261$\pm$9                   & 3.52  $\pm$0.16  $\pm$0.37     & -0.49   $\pm$0.18  $\pm$0.22   & 3.64   $\pm$0.32  $\pm$0.57 \\
      50--60               & 149$\pm$6                   & 6.43  $\pm$0.26  $\pm$0.96     & -1.38   $\pm$0.24  $\pm$0.29   & 6.54   $\pm$0.47  $\pm$0.92 \\
      60--70               & 76$\pm$4                    & 11.91 $\pm$0.53  $\pm$2.1      & -4.90   $\pm$0.58  $\pm$0.56   & 10.34  $\pm$1.0   $\pm$1.8 \\
      70--80               & 35$\pm$2                    & 29.99 $\pm$1.2   $\pm$4.0      & -16.02  $\pm$1.5   $\pm$1.1    & 17.93  $\pm$2.0   $\pm$3.3 \\
      \hline \hline
    \end{tabular}
  }
  \label{tab:nudynresults}
\end{table}
\\ \\
Measured values of \nudynpikashort and \nudynkaprshort are positive across the entire centrality 
range, while \nudynpiprshort is negative for the most peripheral collisions and changes sign at mid-central collisions. 
The centrality dependencies observed in \nudynkaprshort and \nudynpiprshort are similar in shape, being flat from central to mid-central 
collisions and systematically decreasing for the most peripheral ones. In contrast, \nudynpikashort is almost independent of centrality from 
most peripheral to mid-central collisions and rises as the centrality increases. A similar qualitative behaviour is also observed for 
\nudynpikashort within the kinematic range of $|\eta|<1$ and $0.2<p<0.6$~GeV/$c$ as measured in Au--Au collisions at \mbox{\snn$=200$~GeV} 
by the STAR collaboration. The difference in the absolute values is, to a large extent, due to the increase in \dNdeta by almost a factor of 
two between the two collision energies. The same argument holds true for the most central STAR data at \mbox{$62.4$~GeV}, although 
the centrality dependence is rather flat in this case \cite{Abelev:2009ai}. The overall behaviour is defined by the interplay between 
correlation and fluctuation terms encoded in the definition of the \nudyn observable. To disentangle these terms, one needs a dedicated study 
focusing on separate charge combinations, which also makes it possible to investigate contributions from resonance decays and global charge conservations.
\\ \\ 
An important characteristic of HIJING is that it treats nucleus-nucleus collisions as an independent superposition of nucleon-nucleon interactions.
As such, it does not incorporate mechanisms for final-state interactions among the produced particles and therefore phenomena such as 
equilibrium and collectivity do not occur. The AMPT calculations are performed with three different settings including (i) string melting, 
(ii) hadronic rescattering, and (iii) string melting and hadronic rescattering. All three versions of the AMPT model presented 
here use hard minijet partons and soft strings from HIJING as initial conditions. Partonic evolution is described 
by Zhang's parton cascade (ZPC) \cite{Zhang:1997ej} which is followed by a hadronization process. In the last step, hadronic rescattering and the 
decay of resonances takes place. 
In the default AMPT model, after minijet partons stop interacting with other partons, they are combined with their parent
strings to form excited strings, which are then converted to hadrons according to the Lund string fragmentation model \cite{Lin:2004en}. 
However, in the string melting scenario, instead of employing the Lund string fragmentation mechanism, hadronization is modeled via a
quark coalescence scheme by combining two nearest quarks into a meson and three nearest quarks (antiquarks) into a baryon 
(antibaryon). This ultimately reduces the correlation between produced hadrons.
\\ \\
HIJING produces positive values for the three particle pair combinations and does not exhibit any non-monotonic behaviour as a function of centrality, 
even though it implements exact global conservation laws. 
In contrast, hadronic rescattering produces additional resonances at the hadronization phase thereby introducing additional correlations 
between particles \cite{Lin:2004en}. Consequently, the AMPT configuration with hadronic rescattering drives the \nudyn results towards negative values as 
the collision centrality increases. In particular, for \nudynpiprshort, contrary to the data, it predicts negative values.
On the other hand, the AMPT version with string melting shows a weak centrality dependence for the three particle pair combinations.
None of the models investigated in this work give a reasonable quantitative description of the measured data.
\subsection{Energy dependence}
\begin{figure}[h!]
  \centering
  \includegraphics[width=8.5cm]{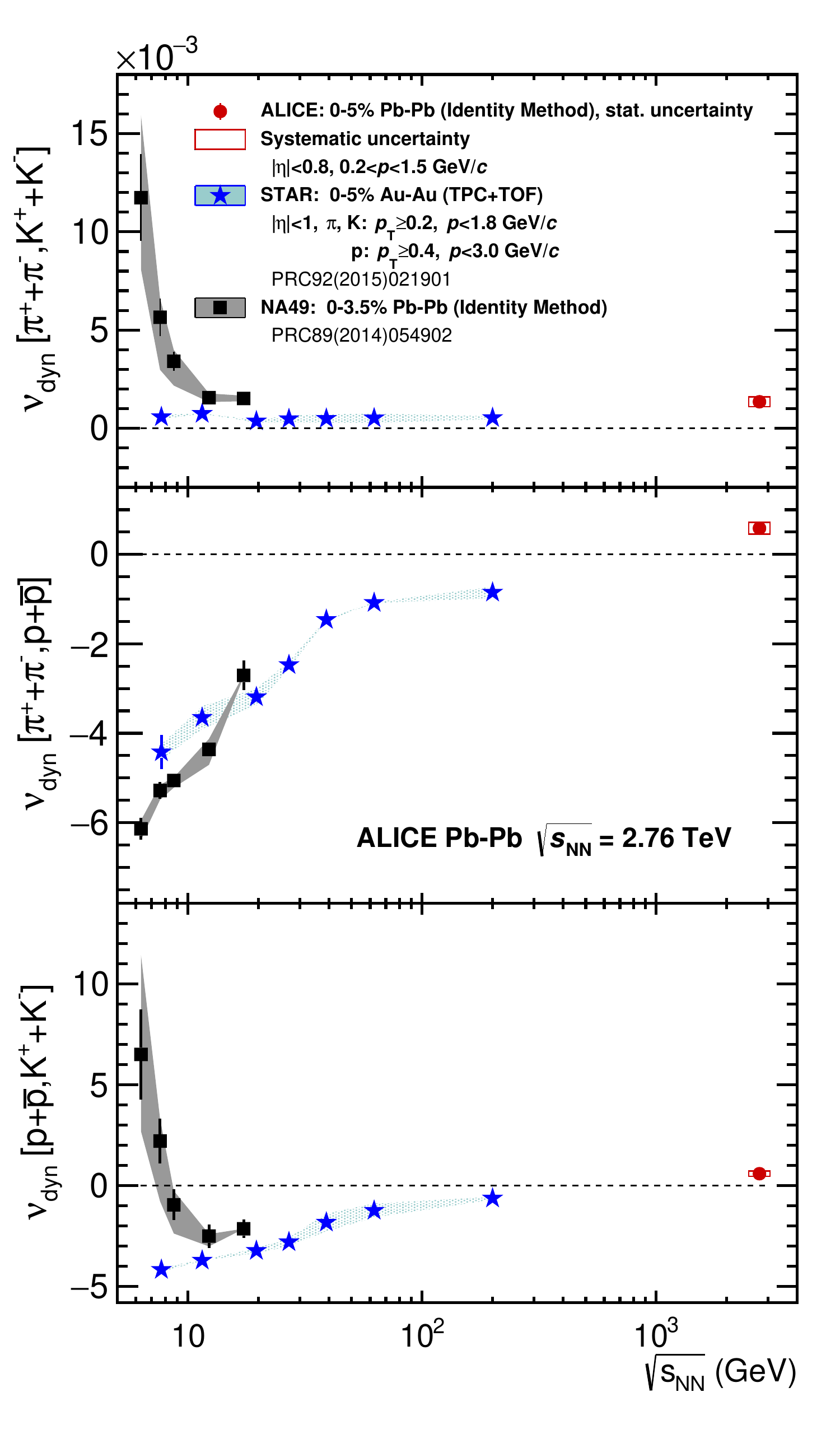}
  \caption{Collision-energy dependence of \nudyn. Results obtained with the Identity Method in this work and by the NA49 collaboration 
  \cite{Anticic:2013htn} in Pb--Pb collisions are shown with red circles and black squares, respectively, 
  while those obtained by the STAR collaboration \cite{Abdelwahab:2014yha} in Au--Au collisions are shown with blue stars.}
  \label{fig:energydep}
\end{figure}

Values of \nudyn measured in this work for the most central Pb--Pb collisions were compared to NA49 and STAR data in \figref{fig:energydep}.
Measurements from NA49 and STAR show a smooth evolution of \nudyn
with collision energy and do not reveal any indications for critical behaviour in the range $6.3 < \sqrt{s_{\rm NN}} < 200$~GeV. 
The apparent differences between NA49 and STAR data for \nudynkaprshort and \nudynpikashort at $\sqrt{s_{\rm NN}} <10$~GeV 
were traced back in Ref~\cite{Anticic:2013htn} to the dependence of \nudyn on the detector acceptance. 
Above this energy, both experiments report positive values for \nudynpikashort, and a weak dependence on the collision energy, whereas 
\nudynkaprshort is negative and approaches zero as the collision energy increases.
\\ \\
ALICE data are positive for the three particle pair combinations and follow the trend observed at lower energies, involving a sign change for 
\nudynpiprshort and \nudynkaprshort as a function of energy. 
Such a change of sign has been predicted by transport models HSD \cite{Gorenstein:2008et} and UrQMD \cite{Bleicher:1999xi} 
in the RHIC energy regime \cite{Abdelwahab:2014yha}. Since neither HSD nor UrQMD explicitly include the quark and gluon degrees of freedom, this observation 
can be attributed to the particular realization of the string and resonance dynamics used in the models \cite{Gorenstein:2008et}.
Additionally, HIJING and AMPT model calculations at LHC energies 
predict positive values except for \nudynpiprshort in the AMPT configuration with hadronic rescattering and without string melting.
To understand the difference between the STAR and ALICE results, the acceptance dependence of \nudyn was also investigated 
with the ALICE data by varying the phase-space coverage. Opening the pseudorapidity window from $|\eta|<0.8$ up to $|\eta|<1$ yields a 
reduction in \nudyn of 10-20\%. 
However, this reduction is insufficient to explain the difference between the ALICE and STAR results, most particularly the sign change with 
increasing energy. 
%
%
\section{\label{sec:summary}Summary}
In summary, measurements of \nudyn in \PbPb collisions at \mbox{\snn$=2.76$~TeV} for three specific particle pair combinations using the Identity Method 
were presented. Values of \nudyn, scaled by the charged-particle density at midrapidity \dNdeta, exhibit finite variations with collision centrality.
This is in contrast to predictions by HIJING, which, for all three pair combinations, show essentially constant as well as positive values. 
The results for \nudynpikashort and \nudynkaprshort are positive across the entire centrality 
range, while \nudynpiprshort changes sign from positive to negative towards more peripheral collisions suggesting differences in the production 
dynamics of these pairs. The centrality dependence of \nudynpikashort shows a similar behaviour, increasing with centrality, as measured in Au--Au collisions at 
\mbox{\snn$=200$~GeV} by the STAR collaboration, while the data at \mbox{\snn$=62.4$~GeV} shows no centrality dependence.
Comparisons with calculations from the AMPT model, using three distinct configurations, show that AMPT is unable to reproduce measured data in 
this work. Calculations with quark coalescence show only a very slight centrality dependence and no sign changes. 
On the other hand, AMPT values with hadronic rescattering and no quark coalescence decrease significantly with increasing collision 
centrality and exhibit a sign change towards central collisions in the case of \nudynpiprshort.
The evolution of \nudyn with collision energy shows that the particle production dynamics changes significantly from that observed at lower energies.
Values of \nudyn measured with all three pair combinations follow a smooth continuation of the data measured by STAR 
and exhibit a change in sign for \nudynkaprshort and \nudynpiprshort.
The analysis of \nudyn with enlarged acceptance shows that the magnitude of \nudyn depends on the kinematical limits
but the change appears too small to explain the difference with the STAR results.
A more detailed analysis of fluctuations with charge and species specific pairs is required to fully characterize the particle 
production dynamics in heavy-ion collisions and understand, in particular, the origin of the sign changes reported in this work. 
%
%
%
\newenvironment{acknowledgement}{\relax}{\relax}
\begin{acknowledgement}
\section*{Acknowledgements}

The ALICE Collaboration would like to thank all its engineers and technicians for their invaluable contributions to the construction of the experiment and the CERN accelerator teams for the outstanding performance of the LHC complex.
The ALICE Collaboration gratefully acknowledges the resources and support provided by all Grid centres and the Worldwide LHC Computing Grid (WLCG) collaboration.
The ALICE Collaboration acknowledges the following funding agencies for their support in building and running the ALICE detector:
A. I. Alikhanyan National Science Laboratory (Yerevan Physics Institute) Foundation (ANSL), State Committee of Science and World Federation of Scientists (WFS), Armenia;
Austrian Academy of Sciences and Nationalstiftung f\"{u}r Forschung, Technologie und Entwicklung, Austria;
Ministry of Communications and High Technologies, National Nuclear Research Center, Azerbaijan;
Conselho Nacional de Desenvolvimento Cient\'{\i}fico e Tecnol\'{o}gico (CNPq), Universidade Federal do Rio Grande do Sul (UFRGS), Financiadora de Estudos e Projetos (Finep) and Funda\c{c}\~{a}o de Amparo \`{a} Pesquisa do Estado de S\~{a}o Paulo (FAPESP), Brazil;
Ministry of Science \& Technology of China (MSTC), National Natural Science Foundation of China (NSFC) and Ministry of Education of China (MOEC) , China;
Ministry of Science, Education and Sport and Croatian Science Foundation, Croatia;
Ministry of Education, Youth and Sports of the Czech Republic, Czech Republic;
The Danish Council for Independent Research | Natural Sciences, the Carlsberg Foundation and Danish National Research Foundation (DNRF), Denmark;
Helsinki Institute of Physics (HIP), Finland;
Commissariat \`{a} l'Energie Atomique (CEA) and Institut National de Physique Nucl\'{e}aire et de Physique des Particules (IN2P3) and Centre National de la Recherche Scientifique (CNRS), France;
Bundesministerium f\"{u}r Bildung, Wissenschaft, Forschung und Technologie (BMBF) and GSI Helmholtzzentrum f\"{u}r Schwerionenforschung GmbH, Germany;
General Secretariat for Research and Technology, Ministry of Education, Research and Religions, Greece;
National Research, Development and Innovation Office, Hungary;
Department of Atomic Energy Government of India (DAE), Department of Science and Technology, Government of India (DST), University Grants Commission, Government of India (UGC) and Council of Scientific and Industrial Research (CSIR), India;
Indonesian Institute of Science, Indonesia;
Centro Fermi - Museo Storico della Fisica e Centro Studi e Ricerche Enrico Fermi and Istituto Nazionale di Fisica Nucleare (INFN), Italy;
Institute for Innovative Science and Technology , Nagasaki Institute of Applied Science (IIST), Japan Society for the Promotion of Science (JSPS) KAKENHI and Japanese Ministry of Education, Culture, Sports, Science and Technology (MEXT), Japan;
Consejo Nacional de Ciencia (CONACYT) y Tecnolog\'{i}a, through Fondo de Cooperaci\'{o}n Internacional en Ciencia y Tecnolog\'{i}a (FONCICYT) and Direcci\'{o}n General de Asuntos del Personal Academico (DGAPA), Mexico;
Nederlandse Organisatie voor Wetenschappelijk Onderzoek (NWO), Netherlands;
The Research Council of Norway, Norway;
Commission on Science and Technology for Sustainable Development in the South (COMSATS), Pakistan;
Pontificia Universidad Cat\'{o}lica del Per\'{u}, Peru;
Ministry of Science and Higher Education and National Science Centre, Poland;
Korea Institute of Science and Technology Information and National Research Foundation of Korea (NRF), Republic of Korea;
Ministry of Education and Scientific Research, Institute of Atomic Physics and Romanian National Agency for Science, Technology and Innovation, Romania;
Joint Institute for Nuclear Research (JINR), Ministry of Education and Science of the Russian Federation and National Research Centre Kurchatov Institute, Russia;
Ministry of Education, Science, Research and Sport of the Slovak Republic, Slovakia;
National Research Foundation of South Africa, South Africa;
Centro de Aplicaciones Tecnol\'{o}gicas y Desarrollo Nuclear (CEADEN), Cubaenerg\'{\i}a, Cuba and Centro de Investigaciones Energ\'{e}ticas, Medioambientales y Tecnol\'{o}gicas (CIEMAT), Spain;
Swedish Research Council (VR) and Knut \& Alice Wallenberg Foundation (KAW), Sweden;
European Organization for Nuclear Research, Switzerland;
National Science and Technology Development Agency (NSDTA), Suranaree University of Technology (SUT) and Office of the Higher Education Commission under NRU project of Thailand, Thailand;
Turkish Atomic Energy Agency (TAEK), Turkey;
National Academy of  Sciences of Ukraine, Ukraine;
Science and Technology Facilities Council (STFC), United Kingdom;
National Science Foundation of the United States of America (NSF) and United States Department of Energy, Office of Nuclear Physics (DOE NP), United States of America.

\end{acknowledgement}
%
%
%
\bibliographystyle{utphys}   
\bibliography{biblio}

\providecommand{\noopsort}[1]{}\providecommand{\singleletter}[1]{#1}%
\providecommand{\href}[2]{#2}\begingroup\raggedright\begin{thebibliography}{10}

\bibitem{Collins:1974ky}
J.~C. Collins and M.~J. Perry, ``{Superdense Matter: Neutrons Or Asymptotically
  Free Quarks?},''
\href{http://dx.doi.org/10.1103/PhysRevLett.34.1353}{{\em Phys. Rev. Lett.}
  {\bfseries 34} (1975) 1353}.

\bibitem{Shuryak:1980tp}
E.~V. Shuryak, ``{Quantum Chromodynamics and the Theory of Superdense
  Matter},''
\href{http://dx.doi.org/10.1016/0370-1573(80)90105-2}{{\em Phys. Rept.}
  {\bfseries 61} (1980) 71--158}.

\bibitem{Stephanov:1998dy}
M.~A. Stephanov, K.~Rajagopal, and E.~V. Shuryak, ``{Signatures of the
  tricritical point in QCD},''
  \href{http://dx.doi.org/10.1103/PhysRevLett.81.4816}{{\em Phys. Rev. Lett.}
  {\bfseries 81} (1998) 4816--4819},
\href{http://arxiv.org/abs/hep-ph/9806219}{{\ttfamily arXiv:hep-ph/9806219
  [hep-ph]}}.

\bibitem{Shuryak:2000pd}
E.~V. Shuryak and M.~A. Stephanov, ``{When can long range charge fluctuations
  serve as a QGP signal?},''
  \href{http://dx.doi.org/10.1103/PhysRevC.63.064903}{{\em Phys. Rev.}
  {\bfseries C63} (2001) 064903},
\href{http://arxiv.org/abs/hep-ph/0010100}{{\ttfamily arXiv:hep-ph/0010100
  [hep-ph]}}.

\bibitem{Bazavov:2012jq}
{\bfseries HotQCD} Collaboration, A.~Bazavov {\em et~al.}, ``{Fluctuations and
  Correlations of net baryon number, electric charge, and strangeness: A
  comparison of lattice QCD results with the hadron resonance gas model},''
  \href{http://dx.doi.org/10.1103/PhysRevD.86.034509}{{\em Phys. Rev.}
  {\bfseries D86} (2012) 034509},
\href{http://arxiv.org/abs/1203.0784}{{\ttfamily arXiv:1203.0784 [hep-lat]}}.

\bibitem{Koch:2008ia}
V.~Koch, \href{http://dx.doi.org/10.1007/978-3-642-01539-7_20}{``{Hadronic
  Fluctuations and Correlations},''} in {\em Relativistic Heavy Ion Physics},
  R.~Stock, ed., pp.~626--652.
\newblock 2010.
\newblock \href{http://arxiv.org/abs/0810.2520}{{\ttfamily arXiv:0810.2520
  [nucl-th]}}.
\newblock
\url{http://materials.springer.com/lb/docs/sm_lbs_978-3-642-01539-7_20}.
\newblock

\bibitem{Bazavov:2011nk}
A.~Bazavov {\em et~al.}, ``{The chiral and deconfinement aspects of the QCD
  transition},'' \href{http://dx.doi.org/10.1103/PhysRevD.85.054503}{{\em Phys.
  Rev.} {\bfseries D85} (2012) 054503},
\href{http://arxiv.org/abs/1111.1710}{{\ttfamily arXiv:1111.1710 [hep-lat]}}.

\bibitem{Pruneau:2002yf}
C.~Pruneau, S.~Gavin, and S.~Voloshin, ``{Methods for the study of particle
  production fluctuations},''
  \href{http://dx.doi.org/10.1103/PhysRevC.66.044904}{{\em Phys. Rev.}
  {\bfseries C66} (2002) 044904},
\href{http://arxiv.org/abs/nucl-ex/0204011}{{\ttfamily arXiv:nucl-ex/0204011
  [nucl-ex]}}.

\bibitem{Christiansen:2009km}
P.~Christiansen, E.~Haslum, and E.~Stenlund, ``{Number-ratio fluctuations in
  high-energy particle production},''
  \href{http://dx.doi.org/10.1103/PhysRevC.80.034903}{{\em Phys. Rev.}
  {\bfseries C80} (2009) 034903},
\href{http://arxiv.org/abs/0902.4788}{{\ttfamily arXiv:0902.4788 [hep-ex]}}.

\bibitem{Abdelwahab:2014yha}
{\bfseries STAR} Collaboration, N.~M. Abdelwahab {\em et~al.}, ``{Energy
  Dependence of $K/\pi$, $p/\pi$, and $K/p$ Fluctuations in Au+Au Collisions
  from $\rm \sqrt{s_{NN}}$ = 7.7 to 200 GeV},''
  \href{http://dx.doi.org/10.1103/PhysRevC.92.021901}{{\em Phys. Rev.}
  {\bfseries C92} no.~2, (2015) 021901},
\href{http://arxiv.org/abs/1410.5375}{{\ttfamily arXiv:1410.5375 [nucl-ex]}}.

\bibitem{Braun-Munzinger:2016yjz}
P.~Braun-Munzinger, A.~Rustamov, and J.~Stachel, ``{Bridging the gap between
  event-by-event fluctuation measurements and theory predictions in
  relativistic nuclear collisions},''
  \href{http://dx.doi.org/10.1016/j.nuclphysa.2017.01.011}{{\em Nucl. Phys.}
  {\bfseries A960} (2017) 114--130},
\href{http://arxiv.org/abs/1612.00702}{{\ttfamily arXiv:1612.00702 [nucl-th]}}.

\bibitem{Abelev:2012pv}
{\bfseries ALICE} Collaboration, B.~Abelev {\em et~al.}, ``{Net-Charge
  Fluctuations in Pb-Pb collisions at $\sqrt{s}_{NN} = 2.76$ TeV},''
  \href{http://dx.doi.org/10.1103/PhysRevLett.110.152301}{{\em Phys. Rev.
  Lett.} {\bfseries 110} no.~15, (2013) 152301},
\href{http://arxiv.org/abs/1207.6068}{{\ttfamily arXiv:1207.6068 [nucl-ex]}}.

\bibitem{Anticic:2013htn}
T.~Anticic {\em et~al.}, ``{Phase-space dependence of particle-ratio
  fluctuations in Pb + Pb collisions from 20 A to 158 A GeV beam energy},''
  \href{http://dx.doi.org/10.1103/PhysRevC.89.054902}{{\em Phys. Rev.}
  {\bfseries C89} no.~5, (2014) 054902},
\href{http://arxiv.org/abs/1310.3428}{{\ttfamily arXiv:1310.3428 [nucl-ex]}}.

\bibitem{Abelev:2014ffa}
{\bfseries ALICE} Collaboration, B.~B. Abelev {\em et~al.}, ``{Performance of
  the ALICE Experiment at the CERN LHC},''
  \href{http://dx.doi.org/10.1142/S0217751X14300440}{{\em Int. J. Mod. Phys.}
  {\bfseries A29} (2014) 1430044},
\href{http://arxiv.org/abs/1402.4476}{{\ttfamily arXiv:1402.4476 [nucl-ex]}}.

\bibitem{Aamodt:2008zz}
{\bfseries ALICE} Collaboration, K.~Aamodt {\em et~al.}, ``{The ALICE
  experiment at the CERN LHC},''
\href{http://dx.doi.org/10.1088/1748-0221/3/08/S08002}{{\em JINST} {\bfseries
  3} (2008) S08002}.

\bibitem{Alver:2008aq}
B.~Alver, M.~Baker, C.~Loizides, and P.~Steinberg, ``{The PHOBOS Glauber Monte
  Carlo},''
\href{http://arxiv.org/abs/0805.4411}{{\ttfamily arXiv:0805.4411 [nucl-ex]}}.

\bibitem{Aamodt:2010jd}
{\bfseries ALICE} Collaboration, K.~Aamodt {\em et~al.}, ``{Suppression of
  Charged Particle Production at Large Transverse Momentum in Central Pb-Pb
  Collisions at $\sqrt{s_{NN}} =$ 2.76 TeV},''
  \href{http://dx.doi.org/10.1016/j.physletb.2010.12.020}{{\em Phys. Lett.}
  {\bfseries B696} (2011) 30--39},
\href{http://arxiv.org/abs/1012.1004}{{\ttfamily arXiv:1012.1004 [nucl-ex]}}.

\bibitem{Abelev:2013vea}
{\bfseries ALICE} Collaboration, B.~Abelev {\em et~al.}, ``{Centrality
  dependence of $\pi$, K, p production in Pb-Pb collisions at $\sqrt{s_{NN}}$ =
  2.76 TeV},'' \href{http://dx.doi.org/10.1103/PhysRevC.88.044910}{{\em Phys.
  Rev.} {\bfseries C88} (2013) 044910},
\href{http://arxiv.org/abs/1303.0737}{{\ttfamily arXiv:1303.0737 [hep-ex]}}.

\bibitem{Gazdzicki:2011xz}
M.~Gazdzicki, K.~Grebieszkow, M.~Mackowiak, and S.~Mrowczynski, ``{Identity
  method to study chemical fluctuations in relativistic heavy-ion
  collisions},'' \href{http://dx.doi.org/10.1103/PhysRevC.83.054907}{{\em Phys.
  Rev.} {\bfseries C83} (2011) 054907},
\href{http://arxiv.org/abs/1103.2887}{{\ttfamily arXiv:1103.2887 [nucl-th]}}.

\bibitem{Gorenstein:2011hr}
M.~I. Gorenstein, ``{Identity Method for Particle Number Fluctuations and
  Correlations},'' \href{http://dx.doi.org/10.1103/PhysRevC.84.024902}{{\em
  Phys. Rev.} {\bfseries C84} (2011) 024902},
\href{http://arxiv.org/abs/1106.4473}{{\ttfamily arXiv:1106.4473 [nucl-th]}}.

\bibitem{Rustamov:2012bx}
A.~Rustamov and M.~I. Gorenstein, ``{Identity Method for Moments of
  Multiplicity Distribution},''
  \href{http://dx.doi.org/10.1103/PhysRevC.86.044906}{{\em Phys. Rev.}
  {\bfseries C86} (2012) 044906},
\href{http://arxiv.org/abs/1204.6632}{{\ttfamily arXiv:1204.6632 [nucl-th]}}.

\bibitem{Gyulassy:1994ew}
M.~Gyulassy and X.-N. Wang, ``{HIJING 1.0: A Monte Carlo program for parton and
  particle production in high-energy hadronic and nuclear collisions},''
  \href{http://dx.doi.org/10.1016/0010-4655(94)90057-4}{{\em Comput. Phys.
  Commun.} {\bfseries 83} (1994) 307},
\href{http://arxiv.org/abs/nucl-th/9502021}{{\ttfamily arXiv:nucl-th/9502021
  [nucl-th]}}.

\bibitem{Deng:2010mv}
W.-T. Deng, X.-N. Wang, and R.~Xu, ``{Hadron production in p+p, p+Pb, and Pb+Pb
  collisions with the HIJING 2.0 model at energies available at the CERN Large
  Hadron Collider},'' \href{http://dx.doi.org/10.1103/PhysRevC.83.014915}{{\em
  Phys. Rev.} {\bfseries C83} (2011) 014915},
\href{http://arxiv.org/abs/1008.1841}{{\ttfamily arXiv:1008.1841 [hep-ph]}}.

\bibitem{Brun:1994aa}
R.~Brun, F.~Bruyant, F.~Carminati, S.~Giani, M.~Maire, A.~McPherson,
  G.~Patrick, and L.~Urban, ``{GEANT Detector Description and Simulation Tool,
  CERN-W5013},'' tech. rep., CERN, 1994.
\newblock \url{http://cds.cern.ch/record/1082634}.

\bibitem{Lin:2004en}
Z.-W. Lin, C.~M. Ko, B.-A. Li, B.~Zhang, and S.~Pal, ``{A Multi-phase transport
  model for relativistic heavy ion collisions},''
  \href{http://dx.doi.org/10.1103/PhysRevC.72.064901}{{\em Phys. Rev.}
  {\bfseries C72} (2005) 064901},
\href{http://arxiv.org/abs/nucl-th/0411110}{{\ttfamily arXiv:nucl-th/0411110
  [nucl-th]}}.

\bibitem{Koch:2009dg}
V.~Koch and T.~Schuster, ``{On the energy dependence of K/pi fluctuations in
  relativistic heavy ion collisions},''
  \href{http://dx.doi.org/10.1103/PhysRevC.81.034910}{{\em Phys. Rev.}
  {\bfseries C81} (2010) 034910},
\href{http://arxiv.org/abs/0911.1160}{{\ttfamily arXiv:0911.1160 [nucl-th]}}.

\bibitem{Abelev:2009ai}
{\bfseries STAR} Collaboration, B.~I. Abelev {\em et~al.}, ``{K/pi Fluctuations
  at Relativistic Energies},''
  \href{http://dx.doi.org/10.1103/PhysRevLett.103.092301}{{\em Phys. Rev.
  Lett.} {\bfseries 103} (2009) 092301},
\href{http://arxiv.org/abs/0901.1795}{{\ttfamily arXiv:0901.1795 [nucl-ex]}}.

\bibitem{Bialas:1976ed}
A.~Bialas, M.~Bleszynski, and W.~Czyz, ``{Multiplicity Distributions in
  Nucleus-Nucleus Collisions at High-Energies},''
\href{http://dx.doi.org/10.1016/0550-3213(76)90329-1}{{\em Nucl. Phys.}
  {\bfseries B111} (1976) 461--476}.

\bibitem{Zhang:1997ej}
B.~Zhang, ``{ZPC 1.0.1: A Parton cascade for ultrarelativistic heavy ion
  collisions},'' \href{http://dx.doi.org/10.1016/S0010-4655(98)00010-1}{{\em
  Comput. Phys. Commun.} {\bfseries 109} (1998) 193--206},
\href{http://arxiv.org/abs/nucl-th/9709009}{{\ttfamily arXiv:nucl-th/9709009
  [nucl-th]}}.

\bibitem{Gorenstein:2008et}
M.~I. Gorenstein, M.~Hauer, V.~P. Konchakovski, and E.~L. Bratkovskaya,
  ``{Fluctuations of the K/pi Ratio in Nucleus-Nucleus Collisions: Statistical
  and Transport Models},''
  \href{http://dx.doi.org/10.1103/PhysRevC.79.024907}{{\em Phys. Rev.}
  {\bfseries C79} (2009) 024907},
\href{http://arxiv.org/abs/0811.3089}{{\ttfamily arXiv:0811.3089 [nucl-th]}}.

\bibitem{Bleicher:1999xi}
M.~Bleicher {\em et~al.}, ``{Relativistic hadron hadron collisions in the
  ultrarelativistic quantum molecular dynamics model},''
  \href{http://dx.doi.org/10.1088/0954-3899/25/9/308}{{\em J. Phys.} {\bfseries
  G25} (1999) 1859--1896},
\href{http://arxiv.org/abs/hep-ph/9909407}{{\ttfamily arXiv:hep-ph/9909407
  [hep-ph]}}.

\end{thebibliography}\endgroup


\providecommand{\noopsort}[1]{}\providecommand{\singleletter}[1]{#1}%
\providecommand{\href}[2]{#2}\begingroup\raggedright\endgroup
%
%
%
\newpage
\appendix
\section{The ALICE Collaboration}
\label{app:collab}

\begingroup
\small
\begin{flushleft}
S.~Acharya\Irefn{org137}\And 
D.~Adamov\'{a}\Irefn{org94}\And 
J.~Adolfsson\Irefn{org34}\And 
M.M.~Aggarwal\Irefn{org99}\And 
G.~Aglieri Rinella\Irefn{org35}\And 
M.~Agnello\Irefn{org31}\And 
N.~Agrawal\Irefn{org48}\And 
Z.~Ahammed\Irefn{org137}\And 
S.U.~Ahn\Irefn{org79}\And 
S.~Aiola\Irefn{org141}
% ...
\And
\renewcommand\labelenumi{\textsuperscript{\theenumi}~}

\section*{Affiliation notes}
\renewcommand\theenumi{\roman{enumi}}
\begin{Authlist}
\item \Adef{org*}Deceased
\item \Adef{orgI}Dipartimento DET del Politecnico di Torino, Turin, Italy
\item \Adef{orgII}M.V. Lomonosov Moscow State University, D.V. Skobeltsyn Institute of Nuclear, Physics, Moscow, Russia
\item \Adef{orgIII}Department of Applied Physics, Aligarh Muslim University, Aligarh, India
\item \Adef{orgIV}Institute of Theoretical Physics, University of Wroclaw, Poland
\end{Authlist}

\section*{Collaboration Institutes}
\renewcommand\theenumi{\arabic{enumi}~}
\begin{Authlist}
\item \Idef{org1}A.I. Alikhanyan National Science Laboratory (Yerevan Physics Institute) Foundation, Yerevan, Armenia
\item \Idef{org2}Benem\'{e}rita Universidad Aut\'{o}noma de Puebla, Puebla, Mexico
\item \Idef{org3}Bogolyubov Institute for Theoretical Physics, Kiev, Ukraine
\item \Idef{org4}Bose Institute, Department of Physics  and Centre for Astroparticle Physics and Space Science (CAPSS), Kolkata, India
\item \Idef{org5}Budker Institute for Nuclear Physics, Novosibirsk, Russia
\item \Idef{org6}California Polytechnic State University, San Luis Obispo, California, United States
\item \Idef{org7}Central China Normal University, Wuhan, China
\item \Idef{org8}Centre de Calcul de l'IN2P3, Villeurbanne, Lyon, France
\item \Idef{org9}Centro de Aplicaciones Tecnol\'{o}gicas y Desarrollo Nuclear (CEADEN), Havana, Cuba
\item \Idef{org10}Centro de Investigaciones Energ\'{e}ticas Medioambientales y Tecnol\'{o}gicas (CIEMAT), Madrid, Spain
\item \Idef{org11}Centro de Investigaci\'{o}n y de Estudios Avanzados (CINVESTAV), Mexico City and M\'{e}rida, Mexico
\item \Idef{org12}Centro Fermi - Museo Storico della Fisica e Centro Studi e Ricerche ``Enrico Fermi', Rome, Italy
\item \Idef{org13}Chicago State University, Chicago, Illinois, United States
\item \Idef{org14}China Institute of Atomic Energy, Beijing, China
\item \Idef{org15}COMSATS Institute of Information Technology (CIIT), Islamabad, Pakistan
\item \Idef{org16}Departamento de F\'{\i}sica de Part\'{\i}culas and IGFAE, Universidad de Santiago de Compostela, Santiago de Compostela, Spain
\item \Idef{org17}Department of Physics, Aligarh Muslim University, Aligarh, India
\item \Idef{org18}Department of Physics, Ohio State University, Columbus, Ohio, United States
\item \Idef{org19}Department of Physics, Pusan National University, Pusan, Republic of Korea
\item \Idef{org20}Department of Physics, Sejong University, Seoul, Republic of Korea
\item \Idef{org21}Department of Physics, University of Oslo, Oslo, Norway
\item \Idef{org22}Department of Physics and Technology, University of Bergen, Bergen, Norway
\item \Idef{org23}Dipartimento di Fisica dell'Universit\`{a} 'La Sapienza' and Sezione INFN, Rome, Italy
\item \Idef{org24}Dipartimento di Fisica dell'Universit\`{a} and Sezione INFN, Cagliari, Italy
\item \Idef{org25}Dipartimento di Fisica dell'Universit\`{a} and Sezione INFN, Trieste, Italy
\item \Idef{org26}Dipartimento di Fisica dell'Universit\`{a} and Sezione INFN, Turin, Italy
\item \Idef{org27}Dipartimento di Fisica e Astronomia dell'Universit\`{a} and Sezione INFN, Bologna, Italy
\item \Idef{org28}Dipartimento di Fisica e Astronomia dell'Universit\`{a} and Sezione INFN, Catania, Italy
\item \Idef{org29}Dipartimento di Fisica e Astronomia dell'Universit\`{a} and Sezione INFN, Padova, Italy
\item \Idef{org30}Dipartimento di Fisica `E.R.~Caianiello' dell'Universit\`{a} and Gruppo Collegato INFN, Salerno, Italy
\item \Idef{org31}Dipartimento DISAT del Politecnico and Sezione INFN, Turin, Italy
\item \Idef{org32}Dipartimento di Scienze e Innovazione Tecnologica dell'Universit\`{a} del Piemonte Orientale and INFN Sezione di Torino, Alessandria, Italy
\item \Idef{org33}Dipartimento Interateneo di Fisica `M.~Merlin' and Sezione INFN, Bari, Italy
\item \Idef{org34}Division of Experimental High Energy Physics, University of Lund, Lund, Sweden
\item \Idef{org35}European Organization for Nuclear Research (CERN), Geneva, Switzerland
\item \Idef{org36}Excellence Cluster Universe, Technische Universit\"{a}t M\"{u}nchen, Munich, Germany
\item \Idef{org37}Faculty of Engineering, Bergen University College, Bergen, Norway
\item \Idef{org38}Faculty of Mathematics, Physics and Informatics, Comenius University, Bratislava, Slovakia
\item \Idef{org39}Faculty of Nuclear Sciences and Physical Engineering, Czech Technical University in Prague, Prague, Czech Republic
\item \Idef{org40}Faculty of Science, P.J.~\v{S}af\'{a}rik University, Ko\v{s}ice, Slovakia
\item \Idef{org41}Faculty of Technology, Buskerud and Vestfold University College, Tonsberg, Norway
\item \Idef{org42}Frankfurt Institute for Advanced Studies, Johann Wolfgang Goethe-Universit\"{a}t Frankfurt, Frankfurt, Germany
\item \Idef{org43}Gangneung-Wonju National University, Gangneung, Republic of Korea
\item \Idef{org44}Gauhati University, Department of Physics, Guwahati, India
\item \Idef{org45}Helmholtz-Institut f\"{u}r Strahlen- und Kernphysik, Rheinische Friedrich-Wilhelms-Universit\"{a}t Bonn, Bonn, Germany
\item \Idef{org46}Helsinki Institute of Physics (HIP), Helsinki, Finland
\item \Idef{org47}Hiroshima University, Hiroshima, Japan
\item \Idef{org48}Indian Institute of Technology Bombay (IIT), Mumbai, India
\item \Idef{org49}Indian Institute of Technology Indore, Indore, India
\item \Idef{org50}Indonesian Institute of Sciences, Jakarta, Indonesia
\item \Idef{org51}INFN, Laboratori Nazionali di Frascati, Frascati, Italy
\item \Idef{org52}INFN, Sezione di Bari, Bari, Italy
\item \Idef{org53}INFN, Sezione di Bologna, Bologna, Italy
\item \Idef{org54}INFN, Sezione di Cagliari, Cagliari, Italy
\item \Idef{org55}INFN, Sezione di Catania, Catania, Italy
\item \Idef{org56}INFN, Sezione di Padova, Padova, Italy
\item \Idef{org57}INFN, Sezione di Roma, Rome, Italy
\item \Idef{org58}INFN, Sezione di Torino, Turin, Italy
\item \Idef{org59}INFN, Sezione di Trieste, Trieste, Italy
\item \Idef{org60}Inha University, Incheon, Republic of Korea
\item \Idef{org61}Institut de Physique Nucl\'eaire d'Orsay (IPNO), Universit\'e Paris-Sud, CNRS-IN2P3, Orsay, France
\item \Idef{org62}Institute for Nuclear Research, Academy of Sciences, Moscow, Russia
\item \Idef{org63}Institute for Subatomic Physics of Utrecht University, Utrecht, Netherlands
\item \Idef{org64}Institute for Theoretical and Experimental Physics, Moscow, Russia
\item \Idef{org65}Institute of Experimental Physics, Slovak Academy of Sciences, Ko\v{s}ice, Slovakia
\item \Idef{org66}Institute of Physics, Academy of Sciences of the Czech Republic, Prague, Czech Republic
\item \Idef{org67}Institute of Physics, Bhubaneswar, India
\item \Idef{org68}Institute of Space Science (ISS), Bucharest, Romania
\item \Idef{org69}Institut f\"{u}r Informatik, Johann Wolfgang Goethe-Universit\"{a}t Frankfurt, Frankfurt, Germany
\item \Idef{org70}Institut f\"{u}r Kernphysik, Johann Wolfgang Goethe-Universit\"{a}t Frankfurt, Frankfurt, Germany
\item \Idef{org71}Institut f\"{u}r Kernphysik, Westf\"{a}lische Wilhelms-Universit\"{a}t M\"{u}nster, M\"{u}nster, Germany
\item \Idef{org72}Instituto de Ciencias Nucleares, Universidad Nacional Aut\'{o}noma de M\'{e}xico, Mexico City, Mexico
\item \Idef{org73}Instituto de F\'{i}sica, Universidade Federal do Rio Grande do Sul (UFRGS), Porto Alegre, Brazil
\item \Idef{org74}Instituto de F\'{\i}sica, Universidad Nacional Aut\'{o}noma de M\'{e}xico, Mexico City, Mexico
\item \Idef{org75}IRFU, CEA, Universit\'{e} Paris-Saclay, Saclay, France
\item \Idef{org76}iThemba LABS, National Research Foundation, Somerset West, South Africa
\item \Idef{org77}Joint Institute for Nuclear Research (JINR), Dubna, Russia
\item \Idef{org78}Konkuk University, Seoul, Republic of Korea
\item \Idef{org79}Korea Institute of Science and Technology Information, Daejeon, Republic of Korea
\item \Idef{org80}KTO Karatay University, Konya, Turkey
\item \Idef{org81}Laboratoire de Physique Subatomique et de Cosmologie, Universit\'{e} Grenoble-Alpes, CNRS-IN2P3, Grenoble, France
\item \Idef{org82}Lawrence Berkeley National Laboratory, Berkeley, California, United States
\item \Idef{org83}Moscow Engineering Physics Institute, Moscow, Russia
\item \Idef{org84}Nagasaki Institute of Applied Science, Nagasaki, Japan
\item \Idef{org85}National and Kapodistrian University of Athens, Physics Department, Athens, Greece
\item \Idef{org86}National Centre for Nuclear Studies, Warsaw, Poland
\item \Idef{org87}National Institute for Physics and Nuclear Engineering, Bucharest, Romania
\item \Idef{org88}National Institute of Science Education and Research, HBNI, Jatni, India
\item \Idef{org89}National Nuclear Research Center, Baku, Azerbaijan
\item \Idef{org90}National Research Centre Kurchatov Institute, Moscow, Russia
\item \Idef{org91}Niels Bohr Institute, University of Copenhagen, Copenhagen, Denmark
\item \Idef{org92}Nikhef, Nationaal instituut voor subatomaire fysica, Amsterdam, Netherlands
\item \Idef{org93}Nuclear Physics Group, STFC Daresbury Laboratory, Daresbury, United Kingdom
\item \Idef{org94}Nuclear Physics Institute, Academy of Sciences of the Czech Republic, \v{R}e\v{z} u Prahy, Czech Republic
\item \Idef{org95}Oak Ridge National Laboratory, Oak Ridge, Tennessee, United States
\item \Idef{org96}Petersburg Nuclear Physics Institute, Gatchina, Russia
\item \Idef{org97}Physics Department, Creighton University, Omaha, Nebraska, United States
\item \Idef{org98}Physics department, Faculty of science, University of Zagreb, Zagreb, Croatia
\item \Idef{org99}Physics Department, Panjab University, Chandigarh, India
\item \Idef{org100}Physics Department, University of Cape Town, Cape Town, South Africa
\item \Idef{org101}Physics Department, University of Jammu, Jammu, India
\item \Idef{org102}Physics Department, University of Rajasthan, Jaipur, India
\item \Idef{org103}Physikalisches Institut, Eberhard Karls Universit\"{a}t T\"{u}bingen, T\"{u}bingen, Germany
\item \Idef{org104}Physikalisches Institut, Ruprecht-Karls-Universit\"{a}t Heidelberg, Heidelberg, Germany
\item \Idef{org105}Physik Department, Technische Universit\"{a}t M\"{u}nchen, Munich, Germany
\item \Idef{org106}Research Division and ExtreMe Matter Institute EMMI, GSI Helmholtzzentrum f\"ur Schwerionenforschung GmbH, Darmstadt, Germany
\item \Idef{org107}Rudjer Bo\v{s}kovi\'{c} Institute, Zagreb, Croatia
\item \Idef{org108}Russian Federal Nuclear Center (VNIIEF), Sarov, Russia
\item \Idef{org109}Saha Institute of Nuclear Physics, Kolkata, India
\item \Idef{org110}School of Physics and Astronomy, University of Birmingham, Birmingham, United Kingdom
\item \Idef{org111}Secci\'{o}n F\'{\i}sica, Departamento de Ciencias, Pontificia Universidad Cat\'{o}lica del Per\'{u}, Lima, Peru
\item \Idef{org112}SSC IHEP of NRC Kurchatov institute, Protvino, Russia
\item \Idef{org113}Stefan Meyer Institut f\"{u}r Subatomare Physik (SMI), Vienna, Austria
\item \Idef{org114}SUBATECH, IMT Atlantique, Universit\'{e} de Nantes, CNRS-IN2P3, Nantes, France
\item \Idef{org115}Suranaree University of Technology, Nakhon Ratchasima, Thailand
\item \Idef{org116}Technical University of Ko\v{s}ice, Ko\v{s}ice, Slovakia
\item \Idef{org117}Technical University of Split FESB, Split, Croatia
\item \Idef{org118}The Henryk Niewodniczanski Institute of Nuclear Physics, Polish Academy of Sciences, Cracow, Poland
\item \Idef{org119}The University of Texas at Austin, Physics Department, Austin, Texas, United States
\item \Idef{org120}Universidad Aut\'{o}noma de Sinaloa, Culiac\'{a}n, Mexico
\item \Idef{org121}Universidade de S\~{a}o Paulo (USP), S\~{a}o Paulo, Brazil
\item \Idef{org122}Universidade Estadual de Campinas (UNICAMP), Campinas, Brazil
\item \Idef{org123}Universidade Federal do ABC, Santo Andre, Brazil
\item \Idef{org124}University of Houston, Houston, Texas, United States
\item \Idef{org125}University of Jyv\"{a}skyl\"{a}, Jyv\"{a}skyl\"{a}, Finland
\item \Idef{org126}University of Liverpool, Liverpool, United Kingdom
\item \Idef{org127}University of Tennessee, Knoxville, Tennessee, United States
\item \Idef{org128}University of the Witwatersrand, Johannesburg, South Africa
\item \Idef{org129}University of Tokyo, Tokyo, Japan
\item \Idef{org130}University of Tsukuba, Tsukuba, Japan
\item \Idef{org131}Universit\'{e} Clermont Auvergne, CNRS/IN2P3, LPC, Clermont-Ferrand, France
\item \Idef{org132}Universit\'{e} de Lyon, Universit\'{e} Lyon 1, CNRS/IN2P3, IPN-Lyon, Villeurbanne, Lyon, France
\item \Idef{org133}Universit\'{e} de Strasbourg, CNRS, IPHC UMR 7178, F-67000 Strasbourg, France, Strasbourg, France
\item \Idef{org134}Universit\`{a} degli Studi di Pavia, Pavia, Italy
\item \Idef{org135}Universit\`{a} di Brescia, Brescia, Italy
\item \Idef{org136}V.~Fock Institute for Physics, St. Petersburg State University, St. Petersburg, Russia
\item \Idef{org137}Variable Energy Cyclotron Centre, Kolkata, India
\item \Idef{org138}Warsaw University of Technology, Warsaw, Poland
\item \Idef{org139}Wayne State University, Detroit, Michigan, United States
\item \Idef{org140}Wigner Research Centre for Physics, Hungarian Academy of Sciences, Budapest, Hungary
\item \Idef{org141}Yale University, New Haven, Connecticut, United States
\item \Idef{org142}Yonsei University, Seoul, Republic of Korea
\item \Idef{org143}Zentrum f\"{u}r Technologietransfer und Telekommunikation (ZTT), Fachhochschule Worms, Worms, Germany
\end{Authlist}
\endgroup
\end{document}